\documentclass[twocolumn]{aastex62}

\received{**}
\revised{**}
\accepted{**}
\submitjournal{ApJ}
\shorttitle{Linear Polarization in H$\alpha$ Flares}
\shortauthors{Kawate \& Hanaoka}

\begin{document}

\title{Infrequent Occurrence of Significant Linear Polarization in H$\alpha$ Solar Flares}

\correspondingauthor{Tomoko Kawate}
\email{kawate@solar.isas.jaxa.jp}

\author{Tomoko Kawate}
\affil{Institute of Space and Astronautical Science, Japan Aerospace Exploration Agency \\
3-1-1 Yoshinodai, Chuo-ku, Sagamihara Kanagawa, 252-5210, JAPAN}

\author{Yoichiro Hanaoka}
\affiliation{National Astronomical Observatory of Japan \\
2-21-1 Osawa, Mitaka, Tokyo 181-8588, JAPAN}

\begin{abstract}
We performed statistical and event studies of linear polarization in the H$\alpha$ line during solar flares.
The statistical study revealed that, among 71 H$\alpha$ flares analyzed, including 64 \textit{GOES} flares, only one event shows significant linear polarization signals.
Such an infrequent occurrence of significant linear polarization in solar flares is consistent with the result by \cite{bian05}, who studied 30 flares and found no polarization signals.
In the event showing the significant polarization, the maximum degree of linear polarization was 1.16$\pm$0.06\%, and the average direction of the polarization deviated by -142.5$\pm$6.0 degrees from the solar north.
The observed polarization degrees and the directions are consistent with the preceding reports  (e.g. \citealp{heno90b,emsl00,hana03}).
These strong linear polarization signals did not appear at major flare ribbons, nor did they correlate with either hard or soft X-ray emissions temporally or spatially.
Instead they appeared at a minor flare kernel, which corresponds to one of the footpoints of a coronal loop. 
The active region caused coronal dimming after the soft X-ray peak.
The observed flare show no direct evidence that the linear polarization is produced by high energy particles, which are often considered to generate the polarization. 
On the other hand, our study suggests the possibility that coronal mass ejections, which have been often observed in flares showing linear polarization signals, play an important role for exciting linear polarization at H$\alpha$ flare kernels.
\end{abstract}


\section{Introduction} 
Polarization in chromospheric lines during solar flares has been examined for a long time.
It was studied in the expectation that polarization measurements of the chromosphere are helpful to diagnose anisotropic velocity distribution of high energy particles accelerated in solar flares (e.g. \citealp{heno90b,emsl00}), especially at the early phase of polarimetric observations during solar flares. 
Later on it was studied under the assumption that the polarization signals demonstrate the dynamics of a magnetic field in the flaring atmosphere (e.g. \citealp{kuck15,klei17,kuri18,anan18}). 
However, the origin of the linear polarization during solar flares is not fully understood yet.

Linear polarization found in flare ribbons, especially in H$\alpha$, has been investigated and reported by many authors.
\cite{heno90b} found a linear polarization signal of 2.5\% in H$\alpha$ in a flare using a half-wave plate with a rotating period of 64~s, and the direction of polarization was toward the disc center. 
They interpreted that the polarization was produced by low-energy protons through the impact polarization, and the protons had anisotropic velocity distribution produced by the transport effects. 
\cite{emsl00} examined a flare showing a linear polarization signal of 2\% using the same technique as \cite{heno90b}, and the signal was directed toward the disc center.
They also analyzed hard X-ray and gamma-ray data, and concluded that accelerated protons with energies $>\sim200$~keV can contain a significant portion of the total energy released during the flare.
They discussed the notion that the energies involved in their unified electron/proton stochastic particle acceleration model are consistent with the proton energy content.
\cite{hana03} found H$\alpha$ linear polarization signals of 1\% using a wave plate with a rotating period of 0.53~s, which lay perpendicular to flare ribbons.
They explained that the origin of the linear polarization can be an anisotropic proton beam.
\cite{firs08} and \cite{firs15} observed a flare which showed linear polarization signals of 4--8\% and 25\% using a double-beam technique; the degrees of polarization depend on the position of the flare ribbons, and the direction of polarization changes by time and position.
The flare showed strong gamma-ray emission produced through the bremsstrahlung of electrons, and the line profile at positions where strong polarization appeared showed the core reversal.
They discussed the possibility that the penetration of high-energy electrons into the dense layers of the chromosphere could lead the impact polarization and a decrease of the H$\alpha$ intensity.

Subsequently to the observations, numerical studies have been carried out to explain the observed degree of polarization based on the dynamics of particles.
For example, \cite{flet98} discussed the linear polarization produced by evaporation-driven flows as being more consistent with the observations than that produced by beams.
\cite{zark00} calculated H$\alpha$ polarization produced by electron-beam impacts into a flaring atmosphere, and concluded that these impacts can produce H$\alpha$ polarization during the early phase of solar flares.
\cite{vogt01} examined hydrogen-proton anisotropic collisions including the effects of polarization in the local radiation field and the energy distribution of high-energy protons.
They concluded that the local radiation field does not increase the H$\alpha$ polarization significantly.

To investigate the origin of the linear polarization in H$\alpha$, statistical investigations have been performed as well.
\cite{bian05} analyzed 30 flares observed with imaging and spectropolarimetry techniques in H$\alpha$ using the Zurich IMaging POLarimeter instrument at the Istituto Ricerche Solari Locarno, which uses a Piezo-elastic modulation technique with modulation speeds of 84 kHz and 42 kHz for linear and circular polarizations, respectively.
They concluded that no flares show such significant linear polarization signals above 0.07\%.
\cite{hana04,hana05} also discussed that only a few flares among dozens of flares show linear polarization, and the polarization does not necessarily appear in big flares.
\cite{firs14} also performed a statistical investigation by the Large Solar Vacuum Telescope, and they found linear polarization of 2--7\% in 13 out of 32 solar flares.
\cite{heno13} discussed possible causes why some authors find linear polarization and others do not; the causes may be lack of spatial and temporal resolutions for dynamic solar flares, cross-talk originated from the atmospheric seeing, and too few data to overcome the lack of the resolutions and errors.

Following the results from the statistical studies, some authors have discussed that the observed polarization are not necessarily excited by collisions of particles.
\cite{step07} calculated the impact polarization produced by protons in solar flares by solving the non-local thermodynamics radiative transfer equations based on a semi-empirical chromosphere model.
They found that the expected degree of linear polarization is below 0.1\% and the tangential resonance-scattering polarization dominates over the impact polarization effect.
\cite{step13} carried out a numerical simulation of two-dimensional radiative transfer, and explained that the linear polarization reaches about 8 \% due to the anisotropy of the radiative field.
\cite{judg15} observed an X1 flare by spectropolarimetry of the \ion{He}{1} 1083~nm line, and calculated non-LTE radiative transfer in thermal slabs. 
They discussed that the linear polarization can be produced by radiative anisotropy arising from the slabs with a certain thickness.

The discrepancy among the observational results are possibly caused by spurious polarization signals, which are produced by slow modulations. 
To investigate the polarization during flares further, a high-speed polarization modulation is required to suppress spurious signals as small as possible.
To clarify the detection probability of the H$\alpha$ linear polarization in flares, we analyzed 71 H$\alpha$ flares including 64 GOES events captured by an identical instrument with reasonably high spatial and temporal resolutions and a high modulation speed.
In this paper, we report the results on statistical and event studies of the H$\alpha$ linear polarization in flaring regions, particularly focusing on the frequency of the observation of the linear polarization in the H$\alpha$ flare kernels.
The paper consists of the following sections.
In section 2, we overview the telescope and polarimeter systems used for the flare observations.
In section 3, we present the analyses and results of the statistical studies of the linear polarization in flares.
In section 4, we further investigate the event that shows a significant linear polarization in flare ribbons.
Finally, in section 5, we summarize our results and discuss the origin of linear polarization during H$\alpha$ flares.

\section{Instruments and Observation Parameters}

Polarization measurements were performed with a ferroelectric liquid crystal polarimeter, which was installed in one of the telescope tubes of the Solar Flare Telescope~\citep{saku95} of the National Astronomical Observatory of Japan. 
The telescope tube had a 20-cm objective lens, and a Zeiss H$\alpha$ Lyot Filter with a passband of 0.25~\AA\ was also installed. 
Data used in our analyses were taken with a DALSA CA-D1-128T camera, whose frame rate is 490~Hz with 128$\times$128~pixels. The spatial sampling was 2.0~arcsec, and the field of view (FOV) was 256$\times$256~arcsec$^2$.

In order to obtain accurate polarization signals in ground-based observations, reduction of seeing-induced errors is crucial.
A typical Fried's seeing parameter in daytime is several centimeter~\citep{kawa11} and the seeing changes typically in several milliseconds.
Thus, a cycle of the polarization modulation should be completed within several milliseconds.
The frequency of the polarization modulation in our observations was about 120~Hz, which makes the seeing effect small.
Four hundred images were recorded in 1~s, and integrated into a single set of four polarization images. 
The cadence of the polarization data was one second, and up to 500 polarization data was stored for each flare event. 
The observations were performed by the same telescope as those reported by \cite{hana03,hana05}, but the data in our study were obtained with an improved polarimeter with higher accuracy~\citep{hana07}.

\begin{figure*}
\plotone{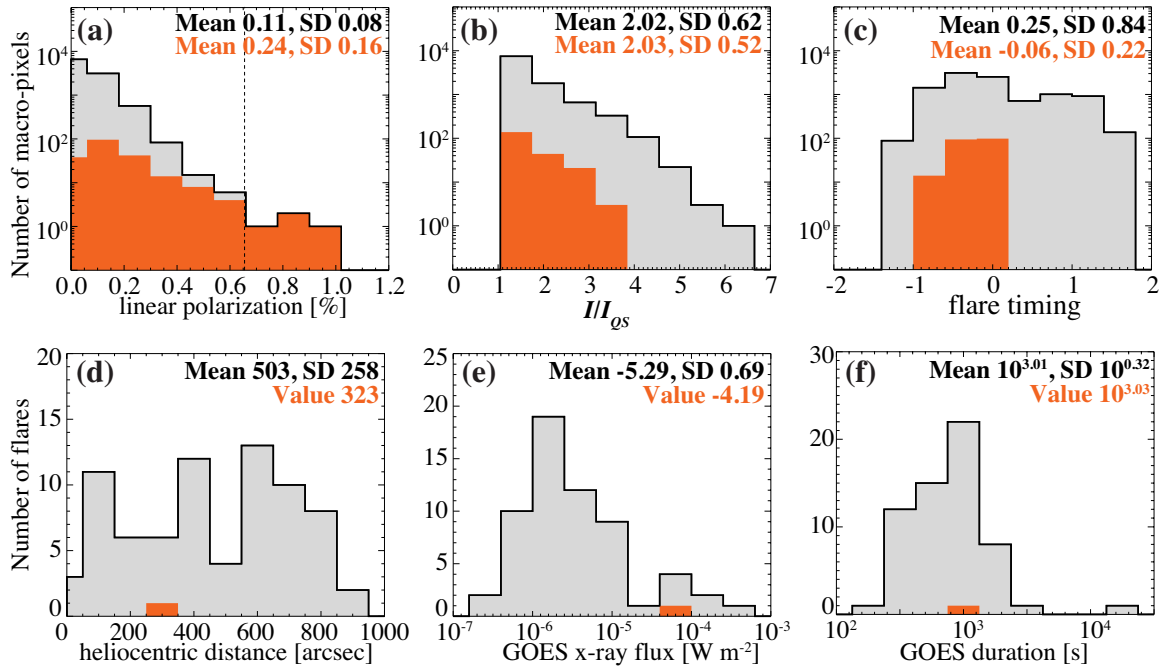}
\caption{Histograms of (a) degree of linear polarization, (b) intensity, (c) observation timing to flare in H$\alpha$ at each macro-pixel. 
Histograms of (d) Heliocentric distance, (e) \textit{GOES} peak X-ray flux, and (g) \textit{GOES} duration of the flares studied. 
In panel (c), the timing is normalized for each flare with setting the \textit{GOES} start time to be $-1$ and \textit{GOES} peak time to be $0$, as described in section 3.1.
The gray histogram in each panel shows the data of the entire events, whereas the red histogram shows the data of the event No. 70. 
Dotted line in (a) shows the sum of 1-$\sigma$ statistical and systematic errors, i.e. 0.65\%.
The average and the standard deviation of the data are denoted in each figure.
 {\label{fig:histo}}}
\end{figure*}

Polarization errors, which consist of statistical and systematic errors, were estimated as follows.
The average of the polarization signals in the quiet region is assumed to be zero at the H$\alpha$ line core.
The measured polarization signals fluctuate due to the statistical uncertainty, and we regarded the standard deviation of the fluctuation as 1-$\sigma$ of the statistical error.
In addition, there are errors due to the incompleteness of the polarimetric calibration as well as errors caused by the seeing effect and the beam deflection during the modulation. 
Polarization patterns caused by these errors change by time and position in the field of view. 
We regarded these errors as systematic errors.

Here, we defined the median brightness of each image as the quiet-sun intensity, $I_{QS}$, and defined the pixels whose intensity is between $0.9I_{QS}$ and $1.1I_{QS}$ as the quiet region. 
On the basis of the polarization signals in such a defined quiet region, the typical 1-$\sigma$ statistical errors were estimated to be 0.4, 0.8, and 0.6\% for Stokes \textit{Q/I}, \textit{U/I}, and \textit{V/I}, respectively, without rebinning or smoothing the pixels.

This instrument was operated regularly between 2004 and 2005.
Flares were automatically detected on the basis of intensity increase at the H$\alpha$ center.
Seventy-one H$\alpha$ flares including 64 \textit{GOES} events were recorded between 2004 July and 2005 December.
The \textit{GOES} events include 4 X-class flares and 13 M-class flares.
The entire list of the events is Appendix A.

In addition to the ground-based data, we used the Extreme ultraviolet Imaging Telescope (EIT; \citealp{dela95}) and the Michelson Doppler Imager (MDI; \citealp{sche95}) onboard the \textit{Solar and Heliospheric Observatory} (\textit{SoHO}; \citealp{domi95}) in order to examine structures of flares.
To understand behavior of thermal and nonthermal particles, we analyzed X-ray data obtained by the \textit{Reuven Ramaty High-Energy Solar Spectroscopic Imager} (\textit{RHESSI}; \citealp{lin02}) and \textit{Geostationary Operational Environmental Satellite} (\textit{GOES}).

\section{Statistical study}
In this section, we compare the polarization and GOES parameters of the 71 flares to examine the characteristics of linear polarization in H$\alpha$ flares.

\subsection{parameter setup}
To reduce the fluctuation, we rebinned the data with the same spatial and temporal samplings as those by \cite{bian05}, namely 10~arcsec and 40~s, respectively.
We call this rebinned pixel a macro-pixel.
With these samplings, the typical statistical error of absolute linear polarization ($\sqrt{Q^2+U^2}/I$) is 0.04\%.
The maximum of the systematic errors in our data is 0.61~\%.

The flaring regions, where we examine the linear polarization, were defined to be the macro-pixels whose intensities are larger than the threshold of $1.5I_{QS}$.
In the 896 images rebinned spatially and temporally, there are 10374 macro-pixels which were defined as the flaring regions.


To specify the relative timing of the acquisition time of each rebinned image, we defined the normalized timing where the \textit{GOES} start and peak times are represented by $-1$ and 0, respectively. For instance, if a strong linear polarization signal is found in the image taken at the \textit{GOES} start time, the timing of the polarization appearance is $-1$.

\subsection{statistical results}
Figures~\ref{fig:histo}(a)--(c) show the histograms of the degree of H$\alpha$ linear polarization, intensity, and relative timing of the 10374 macro-pixels in the flaring regions of the 71 flares.
Figures~\ref{fig:histo}(d)--(f) show the histograms of heliocentric distance of the 71 flares, \textit{GOES} peak X-ray flux for the 64 \textit{GOES} events, and \textit{GOES} flare duration, which is defined by the elapsed time from the \textit{GOES} start to end.
In each panel, we also show the average and the standard deviation of the data.

In Figure~\ref{fig:histo}(a), some macro-pixels show the polarization exceeding the maximum systematic error (0.61\%), and they are presumed to be true polarization signals. 
If we raise the threshold to the sum of the statistical and the systematic errors, i.e. 0.65\%, to avoid picking-up spurious signals more certainly, there are 4 macro-pixels showing the polarization above this value.
There are many macro-pixels of which the degree of polarization is a little below 0.65\%. 
We cannot exclude the possibility that their polarization signals are spurious ones. 
According to Hanaoka (2006), spurious polarization signals with the opposite signs appear adjacently due to slight outfocus.
A careful check of the data revealed that the polarization signals close to the maximum systematic error level are often located at the edge of ribbons and adjacent to another macro-pixel showing the opposite sign of the polarization. 
If the polarization measurements were perfect, these signals cancel each other out.
We regarded that the polarization signals with the opposite signs are not true signals.
On the other hand, all of the four macro-pixels above 0.65\% are observed during the event No. 70 in Table 1, an M6.5 flare. 
Unlike macro-pixels below 0.65\%, they are isolated from those with the polarization of the opposite sign, and they show net polarization. 
Therefore, the detection of the polarization in these macro-pixels is considered to be reliable. 
These facts confirm that the polarization signals of the 4 macro-pixels are true ones, i.e., the solar origin.

\begin{figure}\label{fig:lc}
\plotone{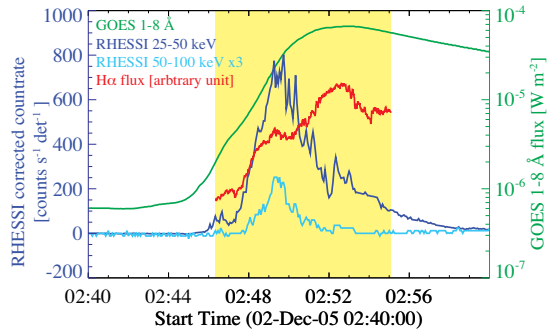}
\caption{Light curves of \textit{GOES} X-ray 1--8~\AA\ flux (green), \textit{RHESSI} 25--50~keV (dark blue) and 50--100~keV (light blue) count rates, and H$\alpha$ flux (red).
The average count rates in the preflare (02:34--02:41~UT) is subtracted from each \textit{RHESSI} light curve.
The yellow region shows the period when the H$\alpha$ high-cadence data are recorded.}\label{fig:lc}
\end{figure}

\begin{figure*}
\plotone{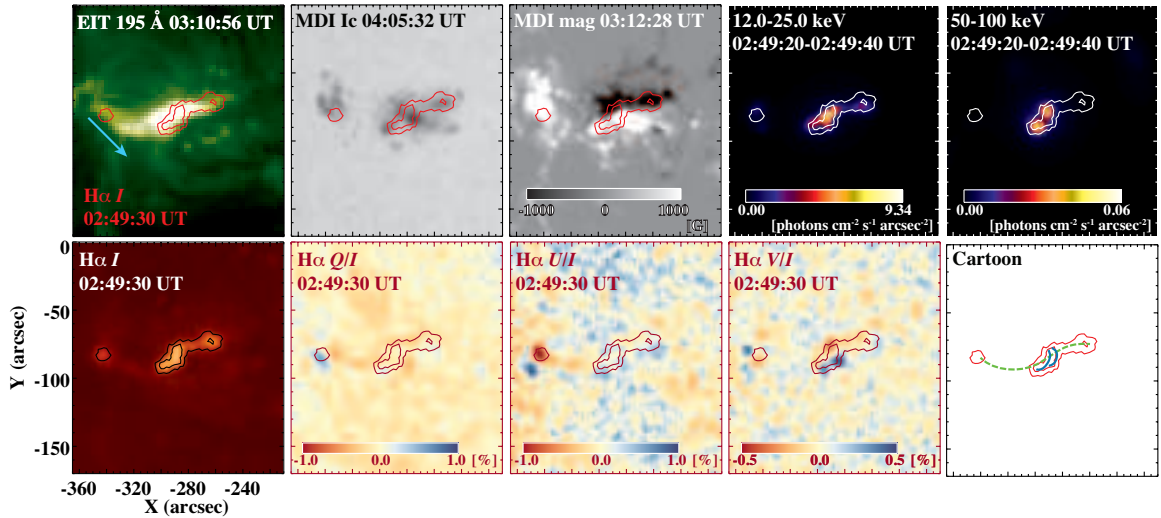}
\caption{Maps of \textit{SoHO}/EIT 195~\AA , \textit{SoHO}/MDI continuum intensity and line-of-sight magnetic field, \textit{RHESSI} 12--25 and 50--100~keV counts, H$\alpha$ Stokes \textit{IQUV}, and cartoon of the coronal magnetic structure drawn in the same field of view. 
In the cartoon, the green dashed line is a coronal loop, and the blue thick lines are small arch structures that connect hard X-ray footpoints.
The contours show H$\alpha$ intensity observed at 02:49:30~UT at the levels of 2$\times$ and 3$\times I_{QS}$. 
The light-blue arrow in the EIT image shows the average direction of the linear polarization in the Eastern H$\alpha$ ribbon during the flare, i.e. $-135.1$ degree from the solar north, as discussed in section 4.2. \label{fig:map}}
\end{figure*}

The details of the large linear polarization signals in event No.~70 will be examined in the next section, but to clarify if this event has unique characteristics other than the polarization, we also show histograms of event No.~70 alone in Figure~\ref{fig:histo}. 
As seen in Figure~\ref{fig:histo}(a), there are also macro-pixels with smaller linear polarization signals in event No.~70.
Figure~\ref{fig:histo}(b) show that the intensity distribution of event No.~70 is (2.03$\pm$0.52)$\times I/I_{QS}$, which are  not different from that of the entire flares, (2.02$\pm$0.62)$\times I/I_{QS}$.
Figure~\ref{fig:histo}(c) shows that the data of event No.~70 were successfully captured between the onset to peak of the flare. 
Figure~\ref{fig:histo}(d) shows that location of event No.~70 is 323 arcsec from the disk center and this is not exceptional among the observed flares, whose locations are widely distributed over the center to the limb.
Figure~\ref{fig:histo}(e) shows that \textit{GOES} X-ray flux of event No.~70 was $6.5\times 10^{-5}$ W{\,}m$^{-2}$, which is relatively strong among the observed data but is not the strongest one ($7.1\times 10^{-4}$ W{\,}m$^{-2}$).
Figure~\ref{fig:histo}(f) shows that the duration of event No.~70 is a typical value among the observed flares, and the event is not a significant long-durational flare as observed by \cite{hana03}. 
Summarizing the above, we cannot find any notable characteristics in event No. 70 except the linear polarization.

\section{Event study}
\begin{figure}
\plotone{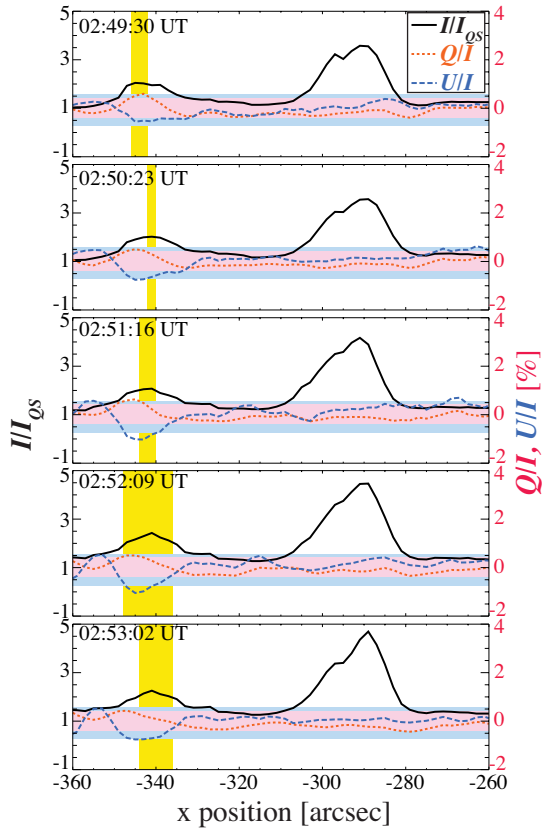}
\caption{Spatial distributions of $I/I_{QS}$ (solid black), $Q/I$ (dotted red), $U/I$ (dashed blue) along the $x$-direction at $y = -87^{\prime\prime}$ at five moments during the flare.
The time is indicated at top left of each panel. 
The pink and light blue belts show the maximum and minimum of systematic errors of $Q/I$ and $U/I$, respectively.
The vertical yellow bar shows the area of $I/I_{QS} >2$, which we define as the flare ribbon in section 4.2. 
\label{fig:kernel}}
\end{figure}

\begin{figure}
\plotone{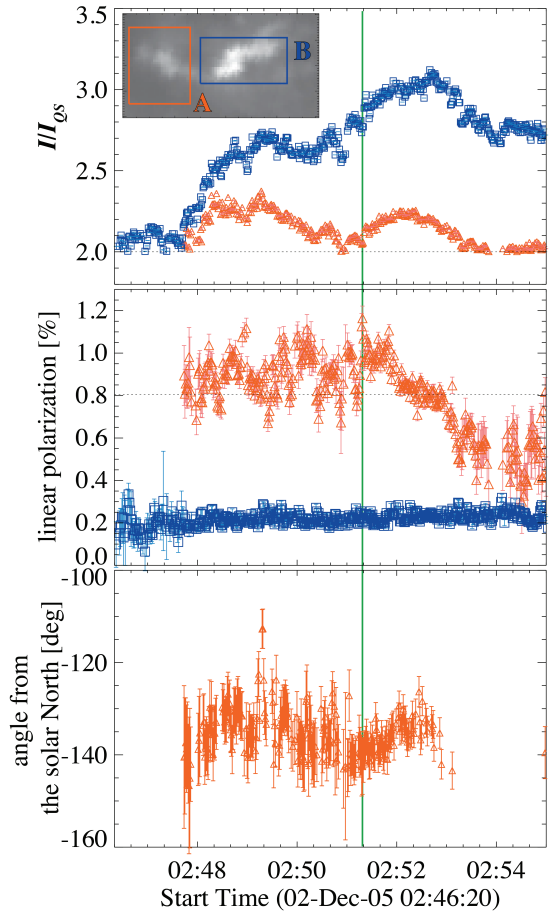}
\caption{Temporal evolution of $I/I_{QS}$, degree of the linear polarization, and direction of the linear polarization.
In the direction of the polarization, we plotted only the data during the appearance of the significant linear polarization signals, and the angle was measured counterclockwise from the solar north.  
The red triangles and the blue squares show data in regions A and B, respectively, which are indicated by red and blue rectangles in the top panel.
The green vertical line indicates the time when the linear polarization degree in region A reached the maximum.
The horizontal dotted line in the middle panel indicates the maximum systematic error of the degree of the linear polarization, i.e., 0.81\%.
\label{fig:pol}}
\end{figure}

\begin{figure}
\plotone{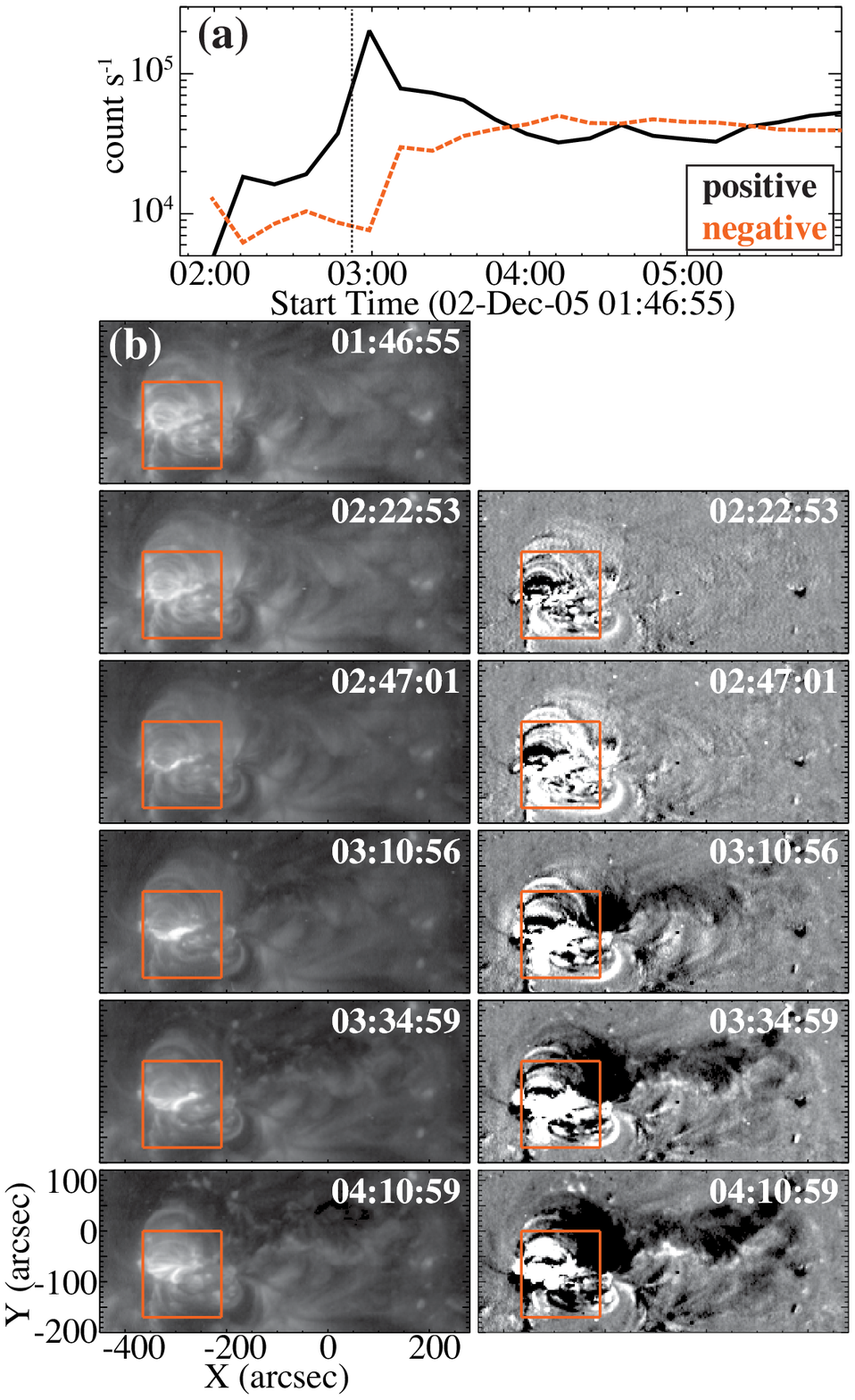}
\caption{(a) Spatially integrated intensity increase (thick black) and decrease (dashed red) of the 195~\AA\ intensity measured on EIS 195~\AA\ images from 01:46:55~UT.
The vertical dotted line indicates the \textit{GOES} peak time at 02:52~UT.
(b) Intensity and base-subtracted maps of EIT 195~\AA . 
The observation time is shown at the top-right of each image.
The base-subtracted images are difference from the image at 01:46:55~UT shown in the first low.
The red boxes show the field of view of Figure~\ref{fig:map}.
\label{fig:dimming}}
\end{figure}

\subsection{overview of the event}
We investigate more details of the event in this section.
The event was an M6.5 class flare occurring in NOAA 10826, located at S03E19.
The \textit{GOES} X-ray flux started to increase at 02:42~UT, peaked at 02:52~UT, and decayed at 03:00~UT. 
\textit{SoHO} and \textit{RHESSI} also observed the flare. 
We show the light curves of \textit{GOES} X-ray flux, \textit{RHESSI} hard X-ray (25--50 and 50--100~keV) counts, and H$\alpha$ flux in Figure~\ref{fig:lc}.
The FOV-integrated H$\alpha$ flux shows two peaks at 02:49:30~UT and 02:52:40~UT, which correspond to the \textit{RHESSI} hard X-ray and \textit{GOES} soft X-ray peaks, respectively.
For this analysis, we used the original temporal resolution of 1~s, but spatially applied a low-pass filter for each image with a cut-off size of 4$^{\prime\prime}$ to suppress the noise.
In these data, the maximum of systematic errors in linear polarization is 0.81\% with the reduced sampling.

\subsection{Spatial distribution of the polarization signals}
First, we examined the position where the strong polarization signals appear during the flare.
Figure~\ref{fig:map} shows Stokes \textit{IQUV} images in H$\alpha$ at the hard X-ray peak time.
The direction of $+Q$ is adopted to be the celestial east-west direction.
The figure also shows co-aligned images of \textit{SoHO}/EIT \ion{Fe}{12} 195~\AA , \textit{SoHO}/MDI continuum intensity and line-of-sight magnetogram, and \textit{RHESSI} 12--25 and 50--100~keV images during the flare.
The \textit{RHESSI} X-ray images are reconstructed by the PIXON algorithm.
Two bright sources appeared in the hard X-ray (50--100~keV) image at ($-300^{\prime\prime}$, $-90^{\prime\prime}$) and ($-290^{\prime\prime}$, $-80^{\prime\prime}$), and they were connected by soft X-ray (12--25~keV) loop-like structures.
The two hard X-ray sources are located in the opposite magnetic polarities seen in the photospheric line-of-sight magnetogram, and this fact suggests that the soft X-ray loops correspond to the magnetic field lines connecting the two hard X-ray sources.
The H$\alpha$ intensity image shows the brightest source corresponding to the X-ray sources, and in addition, two other bright sources located at ($-340^{\prime\prime}$, $-87^{\prime\prime}$) and ($-265^{\prime\prime}$, $-70^{\prime\prime}$).
They are in the opposite magnetic polarities, and connected by a bright EUV loop seen in the 195~\AA\ image.
Thus, we interpreted that small arch structures connect two hard X-ray footpoints and that a loop over the arch structures \citep{moor01} lies approximately along the magnetic neutral line.
We showed a cartoon of the magnetic configuration based on these observations in Figure~\ref{fig:map}.
Among the four H$\alpha$ footpoints, strong \textit{Q/I} and \textit{U/I} signals of $\sim$1\% appeared at ($-340^{\prime\prime}$, $-87^{\prime\prime}$), which we interpreted as a footpoint of the EUV loop.
The averaged degree of the linear polarization at the hard X-ray sources is less than 0.2\%. 
Another source shows an enhancement of 0.95\% outside of the flare brightening at ($-350^{\prime\prime}$, $95^{\prime\prime}$) in the \textit{U/I} map.
The \textit{V/I} map shows an enhancement of $\sim$0.5\% in sunspots, which can be explained by the Zeeman signal emerging due to a slight offset of the transmission wavelength of the Lyot filter from the H$\alpha$ center.
In Figure~\ref{fig:kernel}, we showed spatial distributions of the \textit{IQU} signals along the $x$-axis at $y =-87^{\prime\prime}$, where the strong \textit{Q/I} and \textit{U/I} signals appear, at five moments during the flare.
Peaks of the $Q/I$ and $U/I$ signals are located at 4$^{\prime\prime}$ east from the intensity peak, and size of the region showing significant polarization is 8$\pm$2$^{\prime\prime}$ in the full width at half maximum at 02:52:09~UT.
The maximum absolute values of $Q/I$ and $U/I$ are 0.8$\pm$0.1\% and 1.2$\pm$0.2\%, respectively.
The clear relationship between the macro-pixels with high polarization signals and the solar structure indicates that the the polarization signals are undoubtedly solar origin.

Second, we concentrate on the temporal evolution of two regions where the strong linear polarization appeared (region A) and where the strong hard X-ray emission appeared (region B) individually; the regions are designated in the top panel of Figure~\ref{fig:pol}.
We calculated the absolute degree of the linear polarization, $\sqrt{Q^2+U^2}/I$, and the direction of the linear polarization, $(1/2)\arctan(U/Q)$, at each pixel, and averaged over the pixels with the intensity above $2I/I_{QS}$.
The numbers of pixels that we averaged are up to 52 and 233 in regions A and B, respectively, depending on the observation time.
Figure~\ref{fig:pol} shows the temporal evolutions of pixel-averaged intensities, degree of the linear polarization, and direction of the linear polarization measured from the solar north.
The averaged intensities of the two regions peaked at 02:49:30~UT and 02:52:40~UT, which correspond to the \textit{RHESSI} hard X-ray and \textit{GOES} soft X-ray peaks, respectively.
The linear polarization in region A was above the error level between 02:47:40~UT and 02:53:10~UT, and it reached the maximum value of 1.16$\pm$0.06\% at 02:51:18~UT.
On the other hand, in region B, the degree of the linear polarization was almost a constant value of 0.19\%, which is regarded as the systematic error.
When the degree of the linear polarization in region A reached maximum, the averaged H$\alpha$ intensity and the direction of the linear polarization were $(2.12\pm0.00)I_{QS}$ and $-142.5\pm$6.0 degree from the solar north.
The direction of the linear polarization did not change significantly during the flare, and it stayed around $-135.1$ degree with a standard deviation of 6.1 degree.
In Figure~\ref{fig:map}, we indicate the direction of $-135.1$ degree from the solar north in the EIT image. 
As we see, this direction corresponds the direction of the coronal loop just above the footpoint.
Figures~\ref{fig:map} and \ref{fig:pol} suggest that neither hard X-ray, soft X-ray, nor H$\alpha$ fluxes correlate with the H$\alpha$ linear polarization in region A.

\subsection{Coronal features related to the flare}
Here we examine the coronal structure from ultraviolet observations.
Figure~\ref{fig:dimming} shows \textit{SoHO}/EIT 195~\AA \ intensity images, base-subtracted images, and temporal evolution.
The base-subtracted images show the difference from image taken at 01:46:55~UT.
We integrated positive and negative values over the field of view of the base-subtracted images, and plotted the intensity increase and decrease in Figure~\ref{fig:dimming}(a).
The base-subtracted images show that strong coronal dimming appears in the west of the active region from 03:10~UT, which is 18~min after the \textit{GOES} peak time.
The dimming reached the peak at 04:10~UT, which is 78 minutes after the \textit{GOES} peak time.
Coronal mass ejections (CMEs) are often associated with coronal dimming~\citep{ster97}.
Therefore, in this case, a CME was likely to occur as well, even though the CME catalog\footnote{https://cdaw.gsfc.nasa.gov/CME\_list/} by the SoHO/Large Angle and Spectrometric COronagraph (LASCO; \citealp{brue95}) does not show any CME.

\section{Summary and Discussions} 
In this section, we summarize our results from the statistical and event studies, and discuss the chance to observe the linear polarization in H$\alpha$ and its origin.

\subsection{Infrequent occurrence of significant linear polarization at H$\alpha$ in solar flares}
We found only one event that showed significant linear polarization signals among 71 H$\alpha$ flares including 64 \textit{GOES} flares with the spatial and temporal samplings of $10^{\prime\prime}\times10^{\prime\prime}$ and 40~s, respectively.
Many preceding papers reported the characteristics of the linear polarization in H$\alpha$ only for the events which show significant polarization signals. 
In the cases where the polarization was not detected, it has been considered that the sensitivity of the polarimetry was insufficient.
On the other hand, \cite{bian05} reported that they did not find significant linear polarization signals in 30 flares, even though they used a high-sensitivity instrument.
Based on their preliminary analysis of dozens of flares, \cite{hana04,hana05} also found that one flare showed significant linear polarization signals, while most other flares showed none. 
In our study, we used the polarimeter with higher accuracy described in the papers by \cite{hana04,hana05}, and confirmed their preliminary results. 
Furthermore, although the analyzed samples include some large flares such as 4 X- and 13 M-class events, the results are basically negative.
The high-frequency modulation techniques reduce spurious signals caused by the seeing effect and by flaring structures evolving in time, and data obtained by these techniques are more reliable than  data that were taken by low-frequency modulation techniques.
Thus, our results proved that the linear polarization does not appear in most of the flares, but there are a few flares which actually show the polarization.

The reason \cite{bian05} did not find significant linear polarization is thought to be that the binning size of $10^{\prime\prime}\times10^{\prime\prime}$ and 40~s was too large to resolve the compact flare kernels \citep{heno13,firs14}.
Our results show that the enhancement of the linear polarization extended 8$^{\prime\prime}$ and lasted for 330~s. 
The polarization with such spatial and temporal sizes is still detectable after the binning done by \cite{bian05}.
Our results show that the linear polarization signals cover nearly the entire region of one of the H$\alpha$ footpoints, but are not so high as 10\% or highly localized at the edge of the ribbons.
Nevertheless, due to the limitation of the observation, we cannot conclude that polarization with a smaller scale and a shorter duration does not exist.



\subsection{Origin of the linear polarization at H$\alpha$}
The linear polarization in solar flares observed so far has had the following characteristics:
1) The degree of the polarization is 1--8\% (this study, \citealp{heno90b,emsl00,hana03}).
2) The direction of the polarization is approximately aligned with the coronal loop just above the footpoint (this study, \citealp{hana03}).
3) The degree of the polarization does not correlate with H$\alpha$, soft X-ray, or hard X-ray intensities temporally or spatially (this study).
4) Flares with linear polarization in H$\alpha$ and \ion{He}{1} 10830~\AA\  are often associated with eruptions (this study, \citealp{hana03,xu05,judg15}).
Some of the characteristics are common in the observed flares, and therefore, it is plausible that there is a common mechanism to produce the linear polarization in flares. However, 
from our statistical results, it was concluded that the linear polarization in H$\alpha$ with a few \% is rarely observed. 
The mechanism to produce the H$\alpha$ linear polarization is not a common phenomenon in ordinary flares.
Furthermore, the result that only a minor kernel showed significant polarization means that not all the kernels in a flare necessarily have a common condition relating to the production of the linear polarization.

Let us examine the possibility that the accelerated electrons and the chromospheric dynamics related to the electron injection are the drivers of the linear polarization. They have been considered to produce the linear polarization in many papers.
Nonthermal electrons of $>50$~keV generate Hard X-ray photons of 50--100~keV by bremsstrahlung in the dense chromosphere \citep{koch59}.
Strong electron injection into the chromosphere can induce rapid evaporation \citep{fish85} and return current \citep{emsl80}.
Thus, if anisotropic electrons, return current, or dynamical changes of the chromospheric atmosphere produce linearly polarized photons, linear polarization signals should appear in the same regions as the hard X-ray sources.
If the anisotropic radiative field appearing during chromospheric evaporation is the cause of the linear polarization, the linear polarization signals appear at the edge of the flare ribbons (e.g. \citealp{step13}).
In our results, however, the linear polarization did not correlate with hard X-ray emission. 
In the less bright kernels, the linear polarization source was separated 4$^{\prime\prime}$ from the intensity peak of the kernel but did not concentrate at the edge of the kernel.
Moreover, these mechanisms cannot explain why the significant linear polarization is rarely observed in flares.
Thus, the observed linear polarization cannot be caused by anisotropic electrons or the radiation field discussed above.

Here we focus on another remarkable characteristic of the flares with the linear polarization in H$\alpha$, namely CMEs and eruptions.
From some gamma-ray observations of behind-the-limb flares, protons can be accelerated by CME shocks and propagate toward solar surface along the magnetic field, i.e., toward the footpoints of coronal loops \citep{cliv93,pesc15}. 
Actually, the strong linear polarization appeared at one of the footpoints of a coronal loop in the flare studied here, and the direction of polarization was aligned with the coronal loop just above the footpoint.
Thus, it is possible that the observed polarization was produced through the impact polarization caused by accelerated protons.
Regarding the frequency of observing the proton acceleration, \cite{bai89} showed that 14 flares out of 266 flares were classified as gradual gamma-ray/proton flares by hard X-ray and gamma-ray telescopes onboard the \textit{Solar Maximum Mission}.
Based on the result by \cite{bai89}, the frequency to detect proton acceleration can be 5.3$\pm$1.4\% (or higher, because some weaker events were possibly missed due to the limitation of the instrument).
To produce the impact polarization, accelerated protons should have an anisotropic pitch angle distribution.
Therefore, the probability to detect the impact polarization is less than the probability to detect gamma-ray emission.
This indirect evidence meets the scenario that linear polarization is produced by impact polarization of accelerated protons, but we cannot explain the degree of the polarization of 1\%, as examined by \cite{step07}.
More investigations with highly accurate polarization measurements of solar flares and with numerical simulations of radiative magneto-hydrodynamics at flare kernels in the chromosphere are needed to clarify the origin of linear polarization at H$\alpha$ in solar flares.

\acknowledgments
\textit{SoHO}/LASCO CME catalog is generated and maintained at the CDAW Data Center by NASA and The Catholic University of America in cooperation with the Naval Research Laboratory. 
SOHO is a project of international cooperation between ESA and NASA.
This work was supported by MEXT/JSPS KAKENHI Grant Numbers JP15H05814 and JP17K14314. 
The authors wish to thank Dr. Ichimoto for fruitful discussions.

\facility{GOES, SoHO (EIT), SoHO (MDI), SoHO (LASCO), RHESSI}.
\appendix
\section{Eventlist}

Table~\ref{tab:ev} shows the datasets we analyzed in this paper. 
Note that the datasets include events that were not detected as \textit{GOES} events.

\begin{deluxetable*}{ccccccccc}
\tablecaption{Eventlist\label{tab:ev}}
\tablecolumns{9}
\tablewidth{0pt}
\tablehead{Event No. & \multicolumn{3}{c}{H$\alpha$(UT)} & \multicolumn{3}{c}{\textit{GOES} (UT)}& \textit{GOES} class & position \\
 & start date  & start & end & start & peak & end & & [arcsec] }
\startdata
00 & 2004-07-13 & 00:11:38 & 00:20:24 & 00:09:00 & 00:17:00 & 00:23:00 & M6.7 & ( 649, 181) \\
01 & 2004-07-13 & 05:29:06 & 05:35:14 & 05:21:00 & 05:33:00 & 05:40:00 & C6.7 & ( 692, 184) \\
02 & 2004-07-15 & 01:35:45 & 01:44:32 & 01:30:00 & 01:41:00 & 01:48:00 & X1.8 & (-753,-205) \\
03 & 2004-07-16 & 02:25:36 & 02:29:11 & 01:43:00 & 02:06:00 & 02:12:00 & X1.3 & (-609,-234) \\
04 & 2004-07-17 & 02:02:24 & 02:11:10 & 01:54:00 & 02:01:00 & 02:06:00 & C5.7 & (-407,-246) \\
05 & 2004-07-17 & 02:02:24 & 02:11:10 & 02:07:00 & 02:13:00 & 02:16:00 & C6.8 & (-407,-246) \\
06 & 2004-07-20 & 02:04:51 & 02:13:38 & - & - & - & - & (-609, 4)  \\
07 & 2004-07-20 & 04:12:12 & 04:21:00 & 04:11:00 & 04:23:00 & 04:30:00 & C2.3 & (-594, 3)  \\
08 & 2004-07-20 & 04:55:01 & 05:01:09 & 04:53:00 & 04:57:00 & 05:00:00 & C1.4 & (-581, 2)  \\
09 & 2004-07-20 & 06:19:36 & 06:45:35 & 06:18:00 & 06:22:00 & 06:24:00 & C1.6 & (-509, 81)  \\
10 & 2004-07-22 & 23:35:58 & 23:44:44 & 23:10:00 & 23:24:00 & 23:43:00 & M1.2 & ( 0, -84)  \\
11 & 2004-07-25 & 00:35:37 & 00:44:23 & 00:25:00 & 00:32:00 & 00:36:00 & C7.4 & ( 482, 74)  \\
12 & 2004-07-25 & 03:38:13 & 03:42:50 & - & - & - & - & ( 505, 75)  \\
13 & 2004-07-26 & 23:56:40 & 00:05:26 & 23:46:00 & 00:00:00 & 00:11:00 & M1.2 & ( 754, 112)  \\
14 & 2004-07-27 & 05:42:21 & 05:51:08 & 05:41:00 & 05:45:00 & 05:52:00 & M1.1 & ( 755, -21) \\
15 & 2004-07-28 & 02:56:51 & 06:49:59 & 02:32:00 & 06:09:00 & 08:03:00 & C4.4 & ( 897, 37)  \\
16 & 2004-08-09 & 07:30:28 & 07:38:07 & 07:26:00 & 07:33:00 & 07:39:00 & C2.4 & (-581,-291) \\
17 & 2004-08-12 & 04:59:12 & 05:07:59 & 04:38:00 & 05:05:00 & 05:20:00 & M1.2 & ( 0,-348)  \\
18 & 2004-08-13 & 04:24:13 & 04:32:53 & - & - & - & - & ( 212,-346) \\  
19 & 2004-08-13 & 07:23:06 & 07:31:52 & 06:36:00 & 07:29:00 & 07:38:00 & M1.2 & ( 237,-345)  \\
20 & 2004-08-14 & 04:10:42 & 04:19:29 & 04:10:00 & 04:14:00 & 04:17:00 & M2.4 & ( 462,-304)  \\
21 & 2004-08-18 & 05:07:49 & 05:16:37 & 05:06:00 & 05:10:00 & 05:14:00 & C3.0 & (-303, 92)  \\
22 & 2004-09-08 & 03:42:39 & 03:51:26 & 03:41:00 & 03:49:00 & 03:57:00 & B9.0 & ( -33,-202)  \\
23 & 2004-10-24 & 01:34:35 & 01:43:12 & 01:28:00 & 01:42:00 & 01:44:00 & C1.6 & (-428, 142) \\
24 & 2004-10-25 & 01:51:55 & 01:59:02 & - & - & - & - & (-825, 194) \\
25 & 2004-10-25 & 03:08:32 & 03:17:17 & 02:38:00 & 02:52:00 & 03:03:00 & C4.2 & (-820, 193)  \\
26 & 2004-11-02 & 01:20:51 & 01:27:28 & 01:17:00 & 01:21:00 & 01:23:00 & C1.8 & (-161,-350) \\
27 & 2004-11-04 & 23:58:51 & 00:07:31 & - & - & - & - & (-263, 87) \\
28 & 2004-11-05 & 05:19:20 & 05:28:05 & 05:12:00 & 05:22:00 & 05:25:00 & C2.3 & (-215, 87)  \\
29 & 2004-11-07 & 01:39:12 & 01:47:57 & 01:38:00 & 01:45:00 & 01:51:00 & C3.2 & ( 100, 106)  \\
30 & 2004-11-07 & 04:12:00 & 04:20:46 & 04:11:00 & 04:17:00 & 04:34:00 & C5.3 & ( 133, 107)  \\
31 & 2004-11-09 & 01:05:34 & 01:12:41 & - & - & - & - & ( 652, 92)  \\
32 & 2004-11-09 & 01:39:03 & 01:47:48 & 01:39:00 & 01:43:00 & 01:47:00 & C2.3 & ( 656, 92) \\
33 & 2004-11-09 & 02:51:36 & 02:56:12 & - & - & - & - & ( 664, 93)   \\
34 & 2004-11-10 & 02:03:15 & 02:35:14 & 01:59:00 & 02:13:00 & 02:20:00 & X2.5 & ( 724, 114) \\
35 & 2004-11-16 & 00:10:37 & 00:17:45 & 00:08:00 & 00:13:00 & 00:16:00 & C1.4 & ( 750, 89) \\
36 & 2004-11-16 & 01:41:56 & 01:46:32 & 01:40:00 & 01:45:00 & 01:47:00 & C1.1 & ( 760, 90) \\
37 & 2004-11-16 & 02:23:03 & 02:29:08 & 02:22:00 & 02:25:00 & 02:27:00 & C1.0 & ( 758, 107) \\
38 & 2004-11-27 & 02:00:20 & 02:09:06 & 01:55:00 & 02:04:00 & 02:07:00 & C2.0 & (-599,-301)  \\
39 & 2004-12-01 & 01:39:18 & 01:48:03 & 01:37:00 & 01:45:00 & 01:51:00 & C1.9 & ( 307,-265) \\
40 & 2004-12-02 & 23:45:25 & 00:17:24 & 23:44:00 & 00:06:00 & 00:35:00 & M1.5 & ( 33, 126)  \\
41 & 2004-12-23 & 04:33:02 & 04:41:39 & 04:26:00 & 04:39:00 & 04:51:00 & B8.2 & ( -33,-102)  \\
42 & 2004-12-23 & 05:42:02 & 05:48:02 & 05:41:00 & 05:45:00 & 05:52:00 & B5.1 & ( -33,-102)  \\
43 & 2005-01-14 & 03:50:53 & 04:01:24 & 03:55:00 & 04:04:00 & 04:09:00 & C9.3 & ( -33, -94) \\
44 & 2005-01-17 & 02:51:48 & 03:07:48 & 02:44:00 & 02:59:00 & 03:07:00 & C3.9 & ( 323, 326)  \\
\enddata
\end{deluxetable*}

\setcounter{table}{0}
\begin{deluxetable*}{ccccccccc}
\tablecaption{Eventlist (continued)}
\tablecolumns{9}
\tablewidth{0pt}
\tablehead{Event No. & \multicolumn{3}{c}{H$\alpha$(UT)} & \multicolumn{3}{c}{\textit{GOES} (UT)}& \textit{GOES} class & position \\
 & start date  & start & end & start & peak & end & & [arcsec] }
\startdata
45 & 2005-01-17 & 03:07:49 & 03:24:26 & 03:10:00 & 03:21:00 & 03:32:00 & M2.6 & ( 338, 325)  \\
46 & 2005-01-18 & 02:07:08 & 02:15:45 & 02:06:00 & 02:12:00 & 02:17:00 & C3.2 & ( 439, 355)  \\
47 & 2005-01-19 & 05:13:52 & 05:20:23 & 05:10:00 & 05:26:00 & 05:35:00 & C7.2 & ( 740, 318)  \\
48 & 2005-01-20 & 06:42:57 & 06:51:34 & 06:36:00 & 07:01:00 & 07:26:00 & X7.1 & ( 828, 275) \\
49 & 2005-01-21 & 04:22:34 & 04:31:11 & 04:17:00 & 04:27:00 & 04:34:00 & C6.3 & ( 890, 337) \\
50 & 2005-03-09 & 02:32:09 & 02:40:46 & 02:28:00 & 02:41:00 & 03:00:00 & C1.8 & (-293, 313) \\
51 & 2005-03-18 & 02:35:01 & 02:38:32 & 02:26:00 & 02:35:00 & 02:37:00 & C1.3 & ( 534, -52)  \\
52 & 2005-04-26 & 02:21:24 & 02:30:01 & 02:16:00 & 02:25:00 & 02:30:00 & C1.2 & (-904, -59) \\
53 & 2005-04-26 & 04:45:20 & 04:56:50 & 04:19:00 & 04:41:00 & 04:56:00 & C5.3 & (-898, -57)  \\
54 & 2005-04-28 & 02:26:49 & 02:35:26 & 02:23:00 & 02:36:00 & 02:50:00 & C3.0 & (-684, -31)  \\
55 & 2005-05-15 & 23:53:20 & 00:02:05 & 23:46:00 & 00:03:00 & 00:29:00 & C2.1 & (-313,-224)  \\
56 & 2005-05-16 & 02:51:42 & 03:00:27 & 02:33:00 & 02:43:00 & 02:50:00 & M1.4 & (-266,-239)  \\
57 & 2005-05-16 & 03:32:53 & 03:41:39 & 03:33:00 & 03:36:00 & 03:44:00 & C2.9 & (-159,-222)  \\
58 & 2005-05-17 & 02:33:29 & 02:42:16 & 02:31:00 & 02:39:00 & 02:52:00 & M1.8 & ( 0,-207)  \\
59 & 2005-05-17 & 04:00:32 & 04:02:36 & 03:58:00 & 04:03:00 & 04:05:00 & C8.4 & ( -47,-223) \\
60 & 2005-05-27 & 05:01:52 & 05:10:39 & 05:00:00 & 05:07:00 & 05:17:00 & C2.5 & (-147,-112) \\
61 & 2005-05-27 & 06:24:53 & 06:33:40 & 06:23:00 & 06:35:00 & 06:39:00 & C6.9 & (-131,-112)  \\
62 & 2005-08-03 & 02:32:05 & 02:38:44 & 02:30:00 & 02:36:00 & 02:40:00 & B8.3 & (-194, 69)  \\
63 & 2005-08-03 & 02:44:51 & 02:53:40 & 02:41:00 & 02:45:00 & 02:50:00 & C1.7 & (-161, 101)  \\
64 & 2005-09-17 & 02:08:18 & 02:15:58 & 02:07:00 & 02:12:00 & 02:18:00 & C2.0 & ( 580,-257)  \\
65 & 2005-09-17 & 04:36:47 & 04:45:35 & 04:32:00 & 04:49:00 & 05:08:00 & C3.3 & ( 593,-256)  \\
66 & 2005-09-17 & 06:08:40 & 06:14:21 & 05:58:00 & 06:05:00 & 06:15:00 & M9.8 & ( 593,-256)  \\
67 & 2005-09-23 & 05:51:12 & 05:59:59 & 05:48:00 & 05:59:00 & 06:08:00 & C1.5 & ( -66, 33)  \\
68 & 2005-11-16 & 00:05:48 & 00:14:33 & 00:05:00 & 00:11:00 & 00:20:00 & B5.1 & (-620,-153)  \\
69 & 2005-11-16 & 04:57:29 & 05:03:36 & 04:55:00 & 04:58:00 & 05:01:00 & B2.7 & (-581,-154)  \\
70 & 2005-12-02 & 02:46:20 & 02:55:06 & 02:42:00 & 02:52:00 & 03:00:00 & M6.5 & (-317, -62)  \\
\enddata
\end{deluxetable*}

\clearpage
\bibliographystyle{aasjournal}
\bibliography{s0r_ref}

\end{document}